\documentclass[a4paper]{jpconf}
\usepackage{graphicx}
\usepackage{amssymb}

\font\mybb=msbm10 at 12pt
\def\bb#1{\hbox{\mybb#1}}
\def\bZ {\bb{Z}}
\def\bR {\bb{R}}

\def\bX {\bb{X}}
\def\bP {\bb{P}}

\begin{document}

\begin{flushright}
UMTG-26 \\ DAMTP-2011-103
\end{flushright}

\vskip -1truecm
\title{3D strings and other anyonic things\footnote{Contribution to the XVII  European Workshop on String Theory 2011; Padua, Italy, 5-9 September  2011.} }

\author{Luca Mezincescu$^{\dagger}$ and Paul Townsend${}^\star$}

\address{${}^\dagger\,$Department of Physics, University of Miami, Coral Gables, FL 33124, USA }

\address{${}^\star\,$Department of Applied Mathematical Physics and Theoretical Physics, Centre for Mathematical Sciences, University of Cambridge, Wilberforce Road, Cambridge, CB3 0WA, U.K.}

\ead{mezincescu@server.physics.miami.edu; p.k.townsend@damtp.cam.ac.uk}

\begin{abstract}
We explain how fractional spin and statistics are relevant to  (super)strings in a three-dimensional (3D) Minkowski spacetime.

\end{abstract}

\section*{}
\label{sec:1}
The Nambu-Goto (NG) string,  in a D-dimensional Minkowski spacetime, can be quantized in a unitary gauge that  guarantees the absence of  ghosts, but this can only be done by breaking manifest Lorentz invariance, in which case  Lorentz invariance of the quantum theory is not guaranteed. In fact, as shown long ago by Goddard et al. \cite{Goddard:1973qh}, who quantized the NG string in the light-cone gauge,  the quantum theory breaks Lorentz invariance unless $D=26$. 
Analogous calculations for superstrings lead instead to $D=10$ as the critical dimension. A ghost-free quantization below the critical dimension is possible but only after the introduction of an additional Liouville field on the worldsheet \cite{Polyakov:1981rd} or other  ``longitudinal'' variables \cite{Bardeen:1975gx} that are absent in the critical dimensional.  However, there is an exceptional subcritical dimension for which a unitary gauge quantization preserves Lorentz invariance without the need to introduce any additional variables: $D=3$.  

The usual difficulties with the closure of the Lorentz algebra are trivially absent for $D=3$ \cite{Mezincescu:2010yp,Curtright:2010zz} and a detailed study of closed oriented ``3D'' strings \cite{Mezincescu:2010yp} shows not only that are there no Lorentz anomalies  but  also that there are no supersymmetry anomalies of  the 3D Green-Schwarz (GS) superstrings.  We recall that the classical GS superstring action, which exists for $D=3,4,6,10$ \cite{Green:1983wt},    is an extension of the NG action to describe a string moving in an ${\cal N}=1$ or ${\cal N}=2$  superspace.  Quantization in the critical dimension ($D=10$)  leads to a supersymmetric spectrum of states, with massless ground states, and the same is true for $D=3$. We shall review  the main results to date of light-cone gauge quantization of  3D strings, focusing on closed strings but with some comments on open strings. 

We begin with a brief review of the Hamiltonian formulation of the NG string.  The canonical variables are maps from the string worldsheet, parametrized by $(\tau,\sigma)$ to a Minkowski phase space, with $D$-vector coordinates $(\bX,\bP)$. The action is 
\begin{equation}\label{hamactbos}
S[\bX,\bP;\ell,u] = \int \! d\tau \int_0^{2\pi} \! \frac{d\sigma}{2\pi} 
\left\{\dot {\bX} \cdot {\bP}     
- \frac{1}{2}\ell \left[{\bP}^2 + (T\bX^\prime)^2\right]    - u\,  \bX^\prime \cdot \bP \right\}\, ,  
\end{equation}
where the  coordinate $\sigma$ of the string takes values in the interval $[0,2\pi]$.  If we insist on periodicity in $\sigma$ then we have the action for the closed string; otherwise we have an open string.  
Elimination of the momenta $\bP$ (which can be achieved in the path-integral by Gaussian integration) leads to the  ``Polyakov'' action \cite{Deser:1976rb,Brink:1976sc}:
\begin{equation}\label{Polyakov}
S[\bX,\bP;\ell,u]  \rightarrow S[\bX;g] = -\frac{T}{4\pi} \int \! d^2\xi \,  \sqrt{-\det g}\,  g^{ij} \partial_i \bX \cdot \partial_j\bX\, , 
\end{equation}
where $\xi^i=(\tau,\sigma)$,  and the independent worldsheet metric $g$ is conformally equivalent to the metric 
\begin{equation}
ds^2 =  (u^2 - T^2\ell^2) d\tau^2 + 2u\  d\tau d\sigma +  d\sigma^2
\end{equation}
Any conformal factor drops out of the action due to conformal invariance. It  reappears quantum-mechanically in Polyakov's  covariant path-integral quantization \cite{Polyakov:1981rd}  but is not needed for a  light-cone gauge quantization in the critical  dimension \cite{Goddard:1973qh} and the same is true for $D=3$, at least for free strings. 

The Lagrange multipliers $(\ell,u)$ appearing in the action (\ref{hamactbos}) are analogous to the lapse and shift variables of the Hamiltonian formulation of General Relativity. As in that case, the constraint functions generate gauge invariances that are equivalent to  reparametrization invariance, in this case of the string worldsheet.  In addition to its gauge invariances, the action (\ref{hamactbos}) is  Poincar\'e invariant for  periodic (closed string) or  free-end (open string) boundary conditions. 
In 3D the Poincar\'e  Noether charges are
\begin{equation}\label{noether1}
{\cal P}_\mu = \int_0^{2\pi}\! \frac{d\sigma}{2\pi} \,  {\bP}_\mu \, , \qquad 
{\cal J}^\mu = \int_0^{2\pi}\! \frac{d\sigma}{2\pi}\,  \varepsilon^{\mu\nu\rho}  {\bX}_\nu {\bP}_\rho\, . 
\end{equation}
The two Casimirs of the 3D Poincar\'e group are  
\begin{equation}
{\cal P}^2 =-M^2\, , \qquad {\cal P}\cdot {\cal J} = Ms\, , 
\end{equation} 
and these may be used to classify the unitary irreducible representations \cite{Binegar:1981gv}, in particular $M$ is the mass. The quantity $s$ is sometimes called the ``spin'' but we prefer to reserve this term for the absolute value $|s|$ and call $s$ the ``relativistic helicity'', which we usually abbreviate to ``helicity''.  So defined, spin is unchanged by parity, whereas the helicity changes sign. As both the NG string and its GS superstring extensions preserve parity,  we should expect all massive states of non-zero spin to appear in parity doublets of opposite sign helicity, and it can be shown that this is indeed the case \cite{Mezincescu:2010yp}. 

In our discussion of  3D GS superstrings we shall focus on the ${\cal N}=2$ case  for which the superspace coordinates are $(\bX,\Theta_a)$,  where $\Theta_a$ are a pair ($a=1,2$) of two-component anti-commuting Majorana spinors. The action is obtained in two steps. The first is to replace the differentials $d\bX^\mu$ by the super-translation invariant 1-forms 
$\Pi^\mu = d\bX^\mu + i\bar\Theta_a \Gamma^\mu d\Theta_a$, 
where $\Gamma^\mu$ are the 3D $2\times 2$ Dirac matrices and summation over the index $a$ is implicit. The next step is the addition of a  WZ term constructed from a closed super-Poincar\'e invariant 3-form $h_3$. For $D=3$ there is a 1-parameter family of such 3-forms, which we may parametrize in terms of a unit $2$-vector ${\bf n}$:
\begin{equation}\label{WZ}
h_3({\bf n}) = \Pi^\mu \wedge d\bar\Theta_a \Gamma_\mu \left(n_1 \sigma_3 + n_2 \sigma_1\right)_{ab} \wedge d\Theta_b  \, . 
\end{equation} 
The GS superstring is obtained by choosing ${\bf n}=(1,0)$,  and a coefficient for the resulting WZ term that leads to an additional fermionic gauge invariance. After gauge-fixing, this leads to a cancellation of  zero-point energies and hence massless ground states. 

In the quantum theory each oscillation mode of a string corresponds to some polarization state of a  particle of definite mass-squared. The oscillator 
ground-state energy  of the NG string is negative in the critical dimension, corresponding to a tachyon. It is undetermined in 3D and this complicates a discussion of the spectrum. For this reason, we focus here on the  GS superstrings 
for  which the ground states are those of a massless superparticle.  For the critical dimension this leads to states corresponding to the polarization modes of a $D=10$ linearized supergravity theory. For $D=3$ it  again leads to an irreducible massless supermultiplet. For ${\cal N}=2$ it consists of two bosons and two fermions, which could be identified as the states of a massless vector supermultiplet; this option would introduce a  vector that could couple to objects carrying the central charge allowed by the ${\cal N}=2$ supersymmetry algebra. We say ``option'' because spin is not defined for massless particles in 3D, despite the fact that there is still a distinction between bosons and fermions \cite{Binegar:1981gv,Deser:1991mw}. 

All excited states of the 3D GS superstrings are massive, so let us now look at the structure of massive 3D supermultiplets. We focus on the ${\cal N}=2$ case, for which there are two hermitian spinor charges
${\cal Q}^\alpha_a$, $a=1,2$; the superfix  $\alpha=1,2$ indicates the spinor component. Within a given representation, we may define the new non-hermitian supercharges
\begin{equation}
{\cal S}_a = \frac{1}{\sqrt{2}{\cal P}_-} \left[ \sqrt{2}{\cal P}_- {\cal Q}_a^1 - \left({\cal P}_2 -iM \right){\cal Q}_a^2\right] \, , \qquad {\cal P}_-\equiv \frac{1}{\sqrt{2}} \left({\cal P}_1 - {\cal P}_0\right). 
\end{equation}
The coefficients are particular to the conventions used in \cite{Mezincescu:2010yp}, and in these conventions the anticommutation relations of the ${\cal Q}_a$ supercharges become equivalent to the anticommutation relations
\begin{equation}\label{SQM}
\left\{{\cal S}_a,{\cal S}_b\right\} =0\, , \qquad 
\left\{{\cal S}_a, {\cal S}_b^\dagger \right\} = 2M \left( M \delta_{ab} -i \varepsilon_{ab} Z\right)\, , 
\end{equation}
where we have allowed for a central charge $Z$.  It can be shown that the charges ${\cal S}_a$ lower the helicity $s$ by $1/2$, and that their hermitian conjugates ${\cal S}_a^\dagger$  raise it by $1/2$. Associated with any state annihilated by the ${\cal S}_a$ we then get a supermultiplet by successive actions of the operators ${\cal S}_a^\dagger$ on this state. 

As we explain at the conclusion of this contribution, the possibility of a central charge is relevant to non-perturbative states in the spectrum, assuming that there is an interacting string theory. However, no state in the spectrum of a free ${\cal N}=2$ superstring carries a central charge, so we now set $Z=0$. 
In this case, the construction summarised above leads to four-state supermultiplets with helicities 
\begin{equation}
s= (\bar s-\frac{1}{2}, \bar s, \bar s, \bar s + \frac{1}{2})
\end{equation}
for some average helicity $\bar s$, which is the superhelicity (the value, divided by $M$,  of the Casimir of the super-Poincar\'e group that generalizes $ {\cal P}\cdot {\cal J}$). 
Because parity flips the sign of superhelicity, the unique parity-preserving {\it irreducible} supermultiplet (with $Z=0$) has $\bar s=0$, and hence felicities $s=(-\frac{1}{2},0,0,\frac{1}{2})$. This is just the scalar supermultiplet of 4D ${\cal N}=1$ supersymmetry reduced to 3D.  

One can show that all massive states of the string have at least a 4-fold degeneracy, in addition to the 4-fold degeneracy implied by supersymmetry, so the number of states at each mass level is a multiple of 16. At the first level there are precisely 16 states which appear in 4 copies of the scalar supermultiplet just described. At the next level there are 8 copies of this supermultiplet plus 4 copies of the spin-2 supermultiplet with $\bar s= 3/2$ and 4 copies of the parity-conjugate spin-2 multiplet with $\bar s= -3/2$. Up to this point, there is nothing really unexpected. At the third level we get another 8 copies of the scalar supermultiplet with $\bar s=0$ but the remaining 28+28 supermultiplets, which appear in parity doublets of opposite sign superhelicity,  all have {\it  irrational}  superspin $|\bar s|$, and hence all states in these supermultiplets have  irrational spin. Although the details differ, irrational spins also appear in the spectrum for ${\cal N}=1$ and for the 3D NG string. 

Irrational spins are possible because of a special feature of 3D:  the universal cover of the Lorentz group is not its double cover, the spin group $Sl(2;\bR)$; if it were then  $s$ would be quantized such that $2s \in \bZ$, as in 4D. Instead, the universal cover is an {\it  infinite}  cover  with representations for {\it any} value of $s$. In particular, in any theory with irrational spins the ``Lorentz'' group must be the universal cover of $SO(1,2)$. This has significant implications for quantum statistics via the 3D spin-statistics theorem (see e.g. \cite{Mund:2008hn}) which relates the statistical phase $\theta$ to the helicity $s$ via the formula 
\begin{equation}
e^{i\theta}= e^{2\pi s} \, , 
\end{equation}
so a particle of spin $s$ such that $2s\notin \bZ$ is neither a boson nor a fermion but an   {\it anyon}. Actually, it  would be  preferable to interpret  ``anyon''  to mean  $4s\notin \bZ$ because those cases for which $s=1/4 + n/2$ for integer $n$ are also very special; these are ``semions'' (see e.g. \cite{Sorokin:1992sy}). 
 
The presence of anyons in the spectrum of 3D strings may explain why $D=3$ does not appear to be exceptional  when covariant quantization methods are used, e.g. the Polyakov path-integral approach.
These methods may tacitly assume that the ``Lorentz'' group is $Sl(2;\bR)$. It is worth noting here that covariant quantization is already problematic for a 3D {\it particle} if it is an anyon, so it is 
not surprising that standard approaches to the covariant quantization of strings appear to exclude $D=3$.  One difficulty can be seen from the fact that manifestly Lorentz covariant  anyonic  wave equations are necessarily infinite-dimensional \cite{Jackiw:1990ka,Plyushchay:1990rt}. 

We conclude with a few comments on open 3D strings. In this case each endpoint may be free or 
fixed, or it may be fixed to lie on some curve.  A unitary gauge quantization of the NG string with fixed ends was carried out by Arvis  \cite{Arvis:1983fp}, who concluded that rotational invariance is broken  unless $D=26$. However,  $D=3$ is again exceptional because the relevant rotation group is then trivial,  or abelian if the endpoints coincide. In the case of open GS superstrings,  the boundary conditions must break some of the supersymmetry but there are cases in which half  is preserved. Let us focus on the 
${\cal N}=2$ superstring.   For  free endpoints, the ${\cal N}=2$ supersymmetry is broken to ${\cal N}=1$, as in the critical dimension.  If the endpoints are fixed to lie on some curves, which may be interpreted as D-strings,  then these curves must be parallel straight lines to preserve half the supersymmetry. This possibility is expected in view of the ambiguity noted earlier in the superstring action; 
we may associate the choice ${\bf n}=(0,1)$ in (\ref{WZ}) with the D-string, by analogy to  critical-dimension IIB superstrings (see e.g.  \cite{Bergshoeff:1996tu}). 

The remaining case in which half the  supersymmetry is  preserved occurs when both ends are fixed. If they are fixed at the same point then rotational invariance in the plane is preserved, and this becomes the SO(2) symmetry of a standard supersymmetric quantum mechanics with two supersymmetry charges. We do not know if 3D strings can interact, but if they can it would be natural to interpret the endpoints as D0-branes, which would be described by a centrally-charged ${\cal N}=2$ superparticle with a mass $M$ saturating the Bogomolnyi bound $M\ge |Z|$ that follows from the algebra  (\ref{SQM}). Quantization of this superparticle action leads to a ${\cal N}=2$  supermultiplet in which all states have spin $\frac{1}{4}$  \cite{Mezincescu:2010gb}; in other words, the would-be D0-branes are semions.   If the effective action for a pair of such D0-branes is, in the non-relativistic limit, a supersymmetric $SU(2)$ matrix mechanics with two supersymmetries, as is suggested by critical superstring theory, then the spin-$\frac{1}{4}$ nature of the D0-branes should also be apparent from their statistical phase.  Indeed,  this is precisely the conclusion of Pedder, Sonner and Tong, who considered this model and computed the statistical phase as a Berry phase  \cite{Pedder:2008je}.

\ack

We are grateful to Michael Green, Julian Sonner and David Tong for helpful discussions.  LM acknowleges partial support from National Science Foundation Award 0855386.

\end{document}